\documentclass[ aps,prf,superscriptaddress,amsmath,amssymb,floatfix,preprint,showpacs,showkeys,12pt]{revtex4-1}

\usepackage[utf8]{inputenc}

\usepackage{xcolor}
\usepackage{graphicx}
\usepackage{hyperref}
\usepackage{amssymb}
\usepackage{amsmath}
\usepackage{lineno}
\usepackage{array}
\usepackage{subcaption}
\usepackage{multirow}
\usepackage{csquotes}
\usepackage{amsfonts}
 \usepackage{relsize}
 \usepackage{bbm}
\usepackage{comment}

\begin{document}

\title{Tracking droplets in soft granular flows with deep learning techniques}

\author{Mihir Durve}
\email{mihir.durve@iit.it}
\affiliation{Center for Life Nano- \& Neuro-Science, Fondazione Istituto Italiano di Tecnologia (IIT), 00161 Rome, Italy}

\author{Fabio Bonaccorso}
\affiliation{Center for Life Nano- \& Neuro-Science, Fondazione Istituto Italiano di Tecnologia (IIT), 00161 Rome, Italy}
\affiliation{Istituto per le Applicazioni del Calcolo CNR, via dei Taurini 19, Rome, Italy}
\affiliation{Department of Physics and INFN, University of Rome Tor Vergata, Via della Ricerca Scientifica 1, 00133, Rome, Italy}

\author{Andrea Montessori}
\affiliation{Istituto per le Applicazioni del Calcolo CNR, via dei Taurini 19, Rome, Italy}
\email{a.montessori@iac.cnr.it}

\author{Marco Lauricella}
\affiliation{Istituto per le Applicazioni del Calcolo CNR, via dei Taurini 19, Rome, Italy}

\author{Adriano Tiribocchi}
\affiliation{Istituto per le Applicazioni del Calcolo CNR, via dei Taurini 19, Rome, Italy}

\author{Sauro Succi}
\affiliation{Center for Life Nano- \& Neuro-Science, Fondazione Istituto Italiano di Tecnologia (IIT), 00161 Rome, Italy}
\affiliation{Istituto per le Applicazioni del Calcolo CNR, via dei Taurini 19, Rome, Italy}
\affiliation{Institute for Applied Computational Science, John A. Paulson School of Engineering and Applied Sciences, Harvard University, Cambridge, USA}

\date{\today}

\begin{abstract}
The state-of-the-art deep learning-based object recognition YOLO algorithm and object tracking DeepSORT algorithm are combined to analyze digital images from fluid dynamic simulations of multi-core emulsions and soft flowing crystals and to track moving droplets within these complex flows.
The YOLO network was trained to recognize the droplets with synthetically prepared data, thereby bypassing the labor-intensive data acquisition process.
In both applications, the trained YOLO + DeepSORT procedure performs with high accuracy on the real data from the fluid simulations, with low error levels in the inferred trajectories of the droplets and independently computed ground truth.
Moreover, using commonly used desktop GPUs, the developed application is capable of analyzing data at speeds that exceed the typical image acquisition rates of digital cameras (30 fps), opening the interesting prospect of realizing a low-cost and practical tool to study systems with many moving objects, mostly but not exclusively, biological ones.
Besides its practical applications, the procedure presented here marks the first step towards the automatic extraction of effective equations of motion of many-body soft flowing systems. 
 
\end{abstract}

\maketitle

\section{Introduction}
\label{sec:intro}

Over the last decade, machine learning has taken the scientific world by storm and even more so the commercial one. Even though it is not all old gold that glitters \cite{SucciCoveney}, the potential of machine learning to gain new insights into a variety of complex systems in science and society is unquestionably tantalizing~\cite{Rudin2014,schultz}.

Deep learning is a sub-field of machine learning specifically inspired by the structure and function of the human brain \cite{LeCun2015}. 
Its core consists of artificial neural networks that imitate the operation of the human brain 
in data processing and decision making. 
In the actual digital era,  which generates an unprecedented amount of data (Big Data) 
in the form of images, text, and audio-video clips, deep learning techniques have proved capable of extracting relevant information and correlations~\cite{LeCun2015}, sometimes dramatically 
abating the amount of time required by humans or classical algorithms to perform a similar task. 
%Deep learning techniques are evolving with the digital era, which generates a huge amount of data (called big data) in the form of images, text, audio-video clips, etc. 
%To extract relevant information from the big data, it would take an unprecedented amount of time for humans or classical algorithms to analyze the data. 
%Deep learning networks have shown to be capable of learning relevant features, correlations and extract relevant information from big data~\cite{LeCun2015}. 
This automation in learning from the data has led to the development of many useful computer applications, such as handwriting reading~\cite{darmatasia,ahlawat}, human speech analysis~\cite{tandel,han2014}, text sentiment analysis of posts on social media platforms~\cite{severyn2015,ramadhani2017,zhang2018}, to name a few. In recent years, deep learning networks are being also used to study biological and physical systems~\cite{CARLEO,Alfa_Fold}. In microfluidics, for example, deep learning networks were employed to learn physical parameters, study the size and shapes of the droplets in emulsions~\cite{khor,mahdi}.  

Nowadays, digital cameras assisted by deep learning are used everywhere for traffic management, 
surveillance, crowd management, automated billing, customer services~\cite{osman2017,ho2019, yogameena,ragesh2019}. Given a video feed from these cameras, object detection and subsequent tracking of the detected objects are the two most basic tasks in computer vision before carrying out any further analysis. To achieve these tasks, deep learning algorithms are the fundamental tools. 

%In recent years, multi-layer convolutional neural networks (CNN) have known a surge of popularity for the task of feature extraction for object identification, classification, as well as tracking of the detected objects

In recent years, multi-layer convolutional neural networks (CNN) have a surge of popularity for feature extraction to accomplish object identification, classification, and tracking~\cite{galvez2018, ren2017,wojke,redmon}.  %of the detected objects
The output of the object detection network typically consists of bounding boxes encapsulating the identified objects for visual inspection. 
The tracking network across sequential images then processes the detected objects to construct tracks of moving objects. 
State-of-the-art deep learning models for object detection have demonstrated remarkable accuracy 
compared with classical algorithms and human performance~\cite{redmon,redmon1}.

For the first task of object detection, various deep learning-based models differ in their network architecture, such as the number of convolutional layers, number of pooling layers, filter sizes, and the base algorithm used to analyze the input. At present, Faster region-based convolutional neural networks (FRCNNs)~\cite{ren2017}, You Only Look Once (YOLO)~\cite{redmon}, Single Shot Detector (SSD)~\cite{liu2016} are state of the art in the field. 
Out of these models, we adapt the YOLO algorithm for object detection due to its ease of training and superior image analysis speed, which is further scalable using GPUs. For the second task of tracking the detected objects, we employ a state-of-the-art DeepSORT algorithm~\cite{wojke}, 
which uses the appearance of the objects to track them across a sequence of frames. 
Recently, YOLO and DeepSORT algorithms have been deployed for several real-world applications, such 
as monitoring covid-19 protocols in real-time~\cite{punn2021monitoring,khosravipour}, tracking ball or player trajectories in sports~\cite{zhang2021efficient, host2021}, and many others.

In this work, we develop YOLO + DeepSORT based droplet recognition application 
to analyze the multi-core emulsions and soft granular media simulated via Lattice Boltzmann (LB) methods. 
We train the object detection network to identify droplets from digital images with a synthetically prepared dataset, thereby avoiding the labor-intensive training data gathering process.  In principle, the developed application can also be used to analyze the data generated by the experimental setup of similar physical systems. The goal is to extract trajectories of the center of mass of individual droplets as they move within the flow and then use this information to analyze the dynamics of this complex many-body system. The
application developed in this work can extract trajectories of the droplets in real-time by analyzing the video of the system.

The paper is organized as follows. Before we dive into details of the YOLO + DeepSORT based droplet recognition application, in the next section, we provide a brief account of the Lattice Boltzmann method and relevant simulation details. In section~\ref{sec:technique}, we describe the YOLO and the DeepSORT methods and how the training is implemented. In section~\ref{sec:results}, we report our results on a four-core emulsion and a more complex soft granular material, focusing on the accuracy of the YOLO + DeepSORT procedure in trajectory extraction. Finally, we analyze extracted trajectories of the droplets in a dense emulsion system and compare them with apparently similar active matter systems such as a flock of birds.

\section{Lattice Boltzmann method and simulations}
\label{sec:LB}

We employed two variants of the Lattice Boltzmann approach for the multicomponent flows simulations, namely the color-gradient approach with near-contact interactions and the free energy model for multicomponent fluids.  In Appendices~\ref{app:color_grad} and ~\ref{app:free_energy},  we briefly outline the two models, while in the following sections, we report the simulation details for the two physical systems under consideration.

\subsection{Simulation details for translocation of a soft granular material within narrow channels}

This system is studied using the color gradient LB.
The simulation setup consists of a two-dimensional microfluidic channel made of an inlet reservoir followed by a thinner channel connected to a further downstream reservoir. The height of the chambers is $h=600$ lattice units and the one of the constriction is $h_s=120$ lattice units while its length is $l_s=240$ lattice units. We impose a bounce-back rule for the distribution functions at the walls,
while at the outlet, we employ absorbing (zero gradient) \cite{kruger2017lattice}.

\begin{figure}[h!]
\includegraphics[scale=0.7]{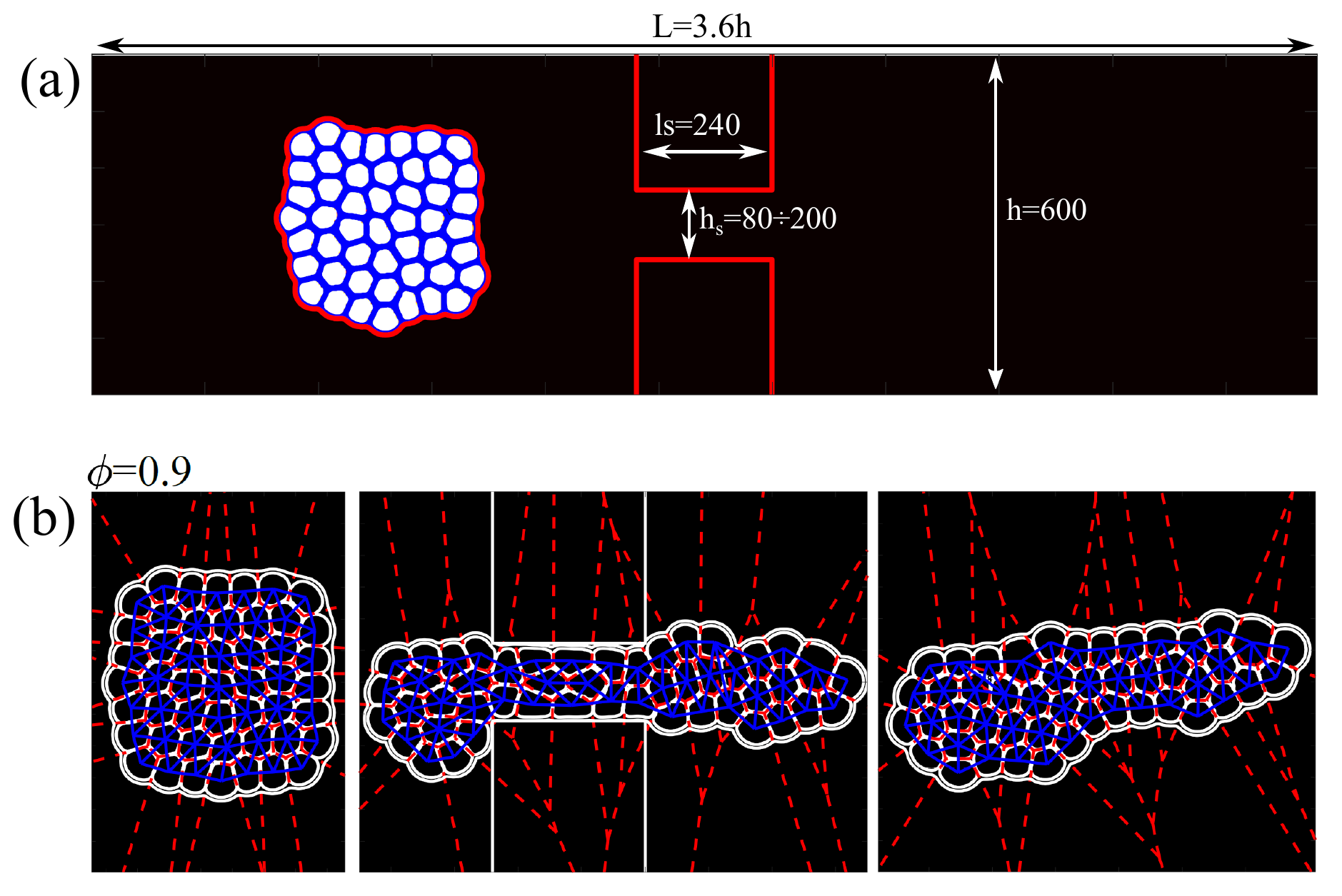}
\caption{\label{fig:translocation} (a) A triple emulsion with $\phi\simeq 0.9$ and $N_d=49$ initialized within the inlet chamber. (b) Translocation dynamics of a dense  $\phi=0.9$, $N_d=49$ and $h_s/D\simeq 0.45$. Once the emulsion crosses the constriction, the resulting shape observed in the outlet chamber crucially depends on the area fraction $\phi$.}
\end{figure}

The soft granular material is formed by droplets (component A, white region in Fig.~\ref{fig:translocation}) immersed within an inter-droplet continuous phase (component B, blue lines) surrounded by an external bulk phase (component C, black region outside the emulsion). Its structure closely resembles that of a high internal phase double emulsion with multi-core morphology \cite{costantini2014,utada2005}. In Fig.~\ref{fig:translocation}(a) we show a snapshot of the domain with the relevant dimension of the geometrical setup and  in Fig.~\ref{fig:translocation}(b) a translocation sequence of the soft material made of $N_d=49$ internal droplets occupying a volume fraction $\phi\simeq 0.9$.
%The emulsion is initialized as a checkerboard-like pattern within the inlet chamber and relaxed for $\sim 5000$ time-steps, enough to achieve a stable configuration.
%In Fig.\ref{fig:translocation} (b) a translocation sequence of the soft material is shown.
In our numerical experiment, a uniform velocity profile of speed  $U_{in}=2\times 10^{-3}$ (in simulation units) is imposed at the inlet, which pushes the emulsion within the narrow constriction. 
Finally, the value of the velocity ensures that Capillary and Reynolds numbers remain well within the typical range of microfluidic experiments ($Ca\sim {\cal O}(10^{-3})$ and $Re\sim {\cal O}(1)$).

\subsection{Simulation details for a flow of a multi-core emulsions within a microchannel.}

Here we report the numerical details of the multi-core emulsion simulated using the free-energy LB. The droplet is set between two flat walls placed at distance $L_y$, where no-slip conditions hold for the velocity field ${\bf v}$ and neutral wetting for the fluid density $\psi$. The former means that the velocity is zero at the boundaries, while the latter that no mass flux crosses the walls
%(${\bf n}\cdot\nabla\mu_i|_{z=0,z=L_z}=0$), where ${\bf n}$ is a unit vector normal to the wall) 
where droplet interfaces are perpendicular. %($\nabla(\nabla^2\phi_i)|_{z=0,z=L_z}=0$).

\begin{figure}[h]
\includegraphics[scale=0.7]{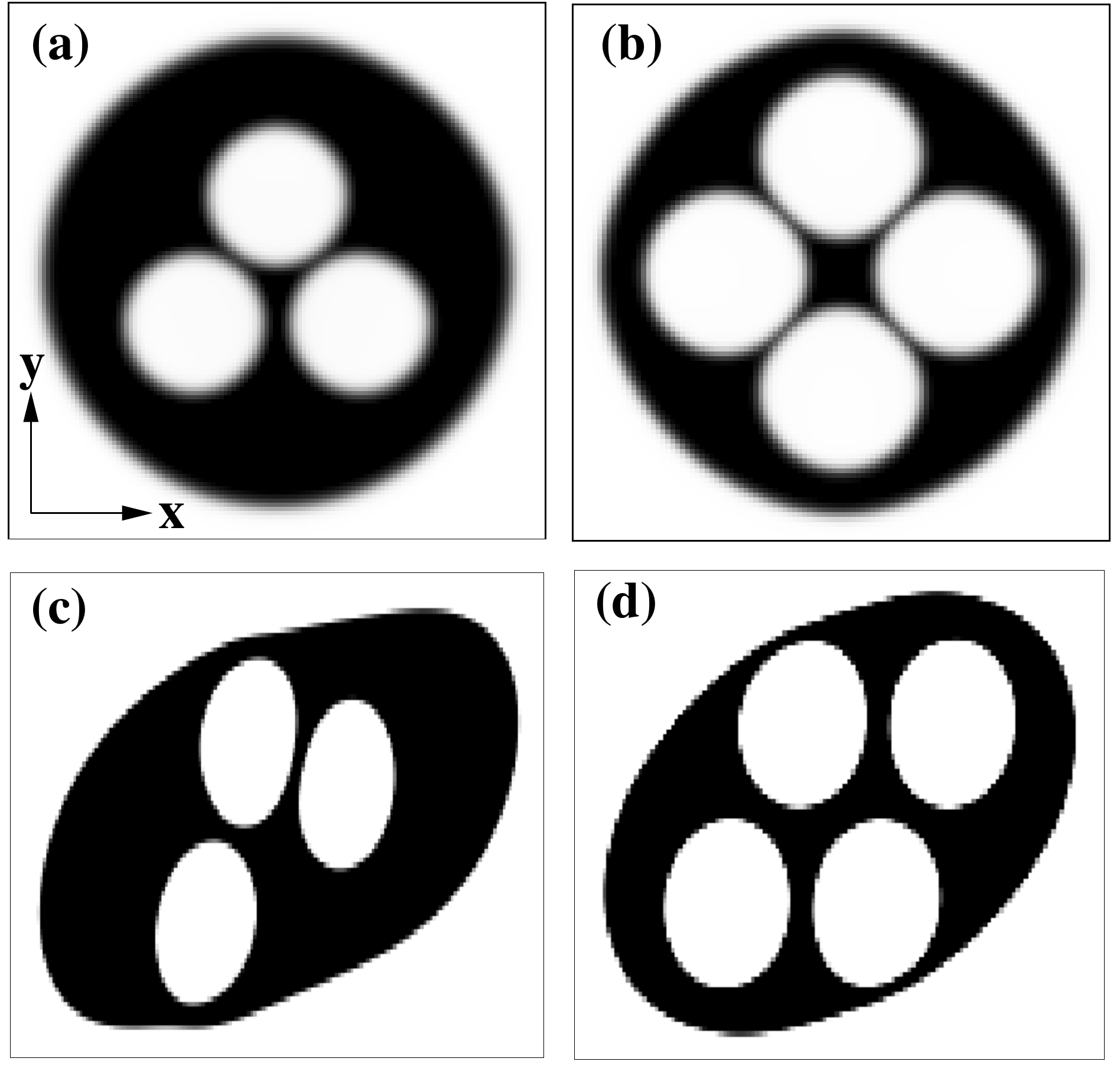}
\caption{ \label{fig:emulsion} (a)-(b) Initial configurations of a three and a four-core monodisperse emulsion. The diameter of the cores is $D_i=30$ lattice sites while that of the shell is $D_O=100$ lattice sites. 
(c)-(d) Examples of instantaneous configurations of a the multi-core emulsions subject to a symmetric shear flow. The shear rate is $\dot{\gamma}\simeq 10^{-4}$ in simulation units. In both cases internal cores rotate periodically clockwise, following an almost elliptical path triggered by the fluid vortex confined within the shell. For more details see Ref.\cite{tiribocchi_pof,tiribocchi_prf}. Here $Ca\simeq 0.2$ and $Re\simeq 1.2$.}
\label{fig_emufree}
\end{figure}

In Fig.\ref{fig_emufree} we show two examples of multi-core emulsions whose design is inspired by experimental realizations. They are made of monodisperse immiscible droplets arranged in a symmetric configuration. In Fig.~\ref{fig_emufree}(a), three cores (white) of diameter $D_i=30$ lattice sites are accommodated within a larger drop (black) of diameter $D_O=100$ lattice sites, in turn, surrounded by a further fluid (white). An analogous setup is used for the four-core emulsion (See Fig.~\ref{fig_emufree}(b)). Starting from these configurations, the emulsions are then driven out of equilibrium by applying a symmetric shear flow, where the top wall moves rightwards (along the positive $x$ axis) at constant speed $v_w$ and the bottom wall in the opposite direction at speed $-v_w$. This sets a shear rate of $\dot{\gamma}=2L_y/v_w$. In previous works \cite{tiribocchi_pof,tiribocchi_prf}, it has been shown that, once the shear is turned on, the external shell elongates and aligns along the direction imposed by the flow, while the internal cores acquire a periodic rotational motion triggered by a fluid vortex formed within the emulsion. These dynamics persist over long periods of time provided that the shear is on. In Fig.~\ref{fig_emufree}(c) and Fig.~\ref{fig_emufree}(d) we show two instantaneous configurations of a three and a four-core emulsion at late times, where $\dot{\gamma}\simeq 10^{-4}$ with $v_w=0.01$ and $L_y=170$ in simulation units.
This system approximately corresponds to a real multi-core emulsion with a shell of diameter $\simeq100\mu$m and cores of diameter $\simeq 30\mu$m, having a surface tension ranging between $1-10$mN/m, a viscosity $\simeq 10^{-1}$ Pa$\cdot$s and a shear rates varying between $0.1-1$/s. Like in the previous method, Capillary and Reynolds numbers here range between $0.1$ and $1$ as well, as in typical microfluidic experiments.

The output of the simulations of the physical systems described above are saved as video files and later given as input to the deep learning-based application for automatic extraction of the trajectories of the moving droplets. In the supplementary material,  see video1.avi file for the simulation output of translocation of triple emulsions within narrow channels and video2.avi for a flow of multi-core emulsions within a microchannel.

\section{Technique}
\label{sec:technique}

\begin{figure}[h]
\includegraphics[width= \textwidth, height=15cm,keepaspectratio]{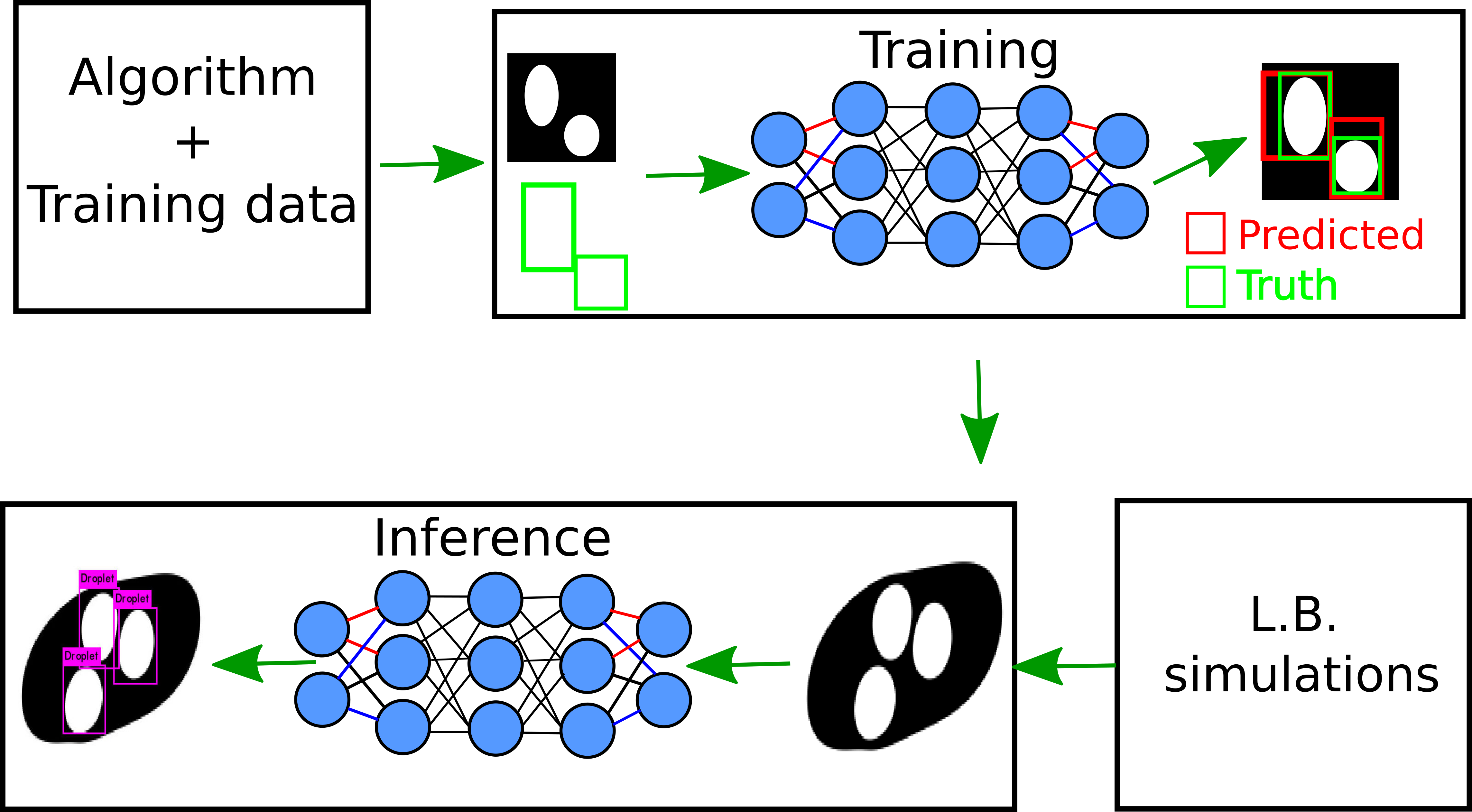}
\caption{\label{fig:Illustration} Typical steps in developing the deep learning-based application for object detection. In the training phase, difference between prediction and truth is used to improve the accuracy of the network by updating the parameters of the network. At the inference stage, a real world data, in our case from LB simulation, is fed to the network to obtain the bounding boxes around the droplets.}
\end{figure}

In this work, we develop an application to extract trajectories of individual droplets simulated via LB methods by adapting two different algorithms, one for object detection and another one for object tracking. 
The first task consists of training an artificial neural network to analyze digital images and locate the objects of interest, which in this case are the droplets. Later, the second task is to track the located objects across multiple sequential frames to infer their trajectories. Fig.~\ref{fig:Illustration} shows the typical steps involved to accomplish the object recognition task~\cite{durve2021}. In the following section, we briefly describe the algorithm employed for the object recognition. 

\subsection{You Only Look Once (YOLO)}

You Only Look Once (YOLO) is a state-of-the-art algorithm 
%used for object detection and classification 
that employs a single multi-layer network for object identification and classification. 
The YOLO algorithm has shown remarkable accuracy with high processing speed in identifying 
and classifying objects from the COCO dataset, which consists of 80 types (classes) of everyday objects~\cite{redmon}.

\begin{figure}[!h]
\includegraphics[width= \textwidth, height=15cm,keepaspectratio]{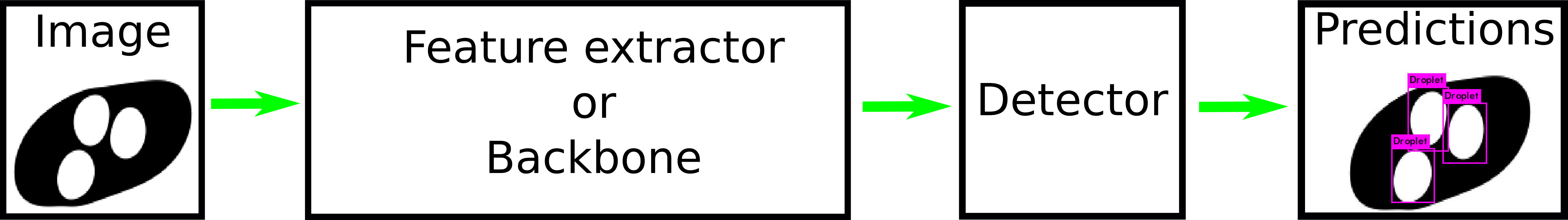}
\caption{\label{yolo_structure} High level sketch of the YOLO network structure.}
\end{figure}

The building structure of the YOLO network consists of two main components, a feature extractor and a detector. 
When an image is given as an input to the YOLO network, the extractor (also called backbone) 
produces feature representations at different scales (see Fig.~\ref{yolo_structure}). 
Such representations are then passed to the detector that outputs the bounding boxes 
along with the confidence score and the classes of the detected objects. 
In this work, we use the YOLO algorithm to identify droplets in the digital images produced by LB simulations of two physical systems. We employ the YOLO-v3 network, which uses the Darknet-53 network as a 
backbone. The Darknet-53 is a deep network consisting of 53 layers, which has shown significant improvement in object detection over its predecessors YOLO-v2 and YOLO-v1, which employ backbone with fewer layers~\cite{redmon1}. 

\begin{figure}[!h]
\includegraphics[width=13cm, height=15cm,keepaspectratio]{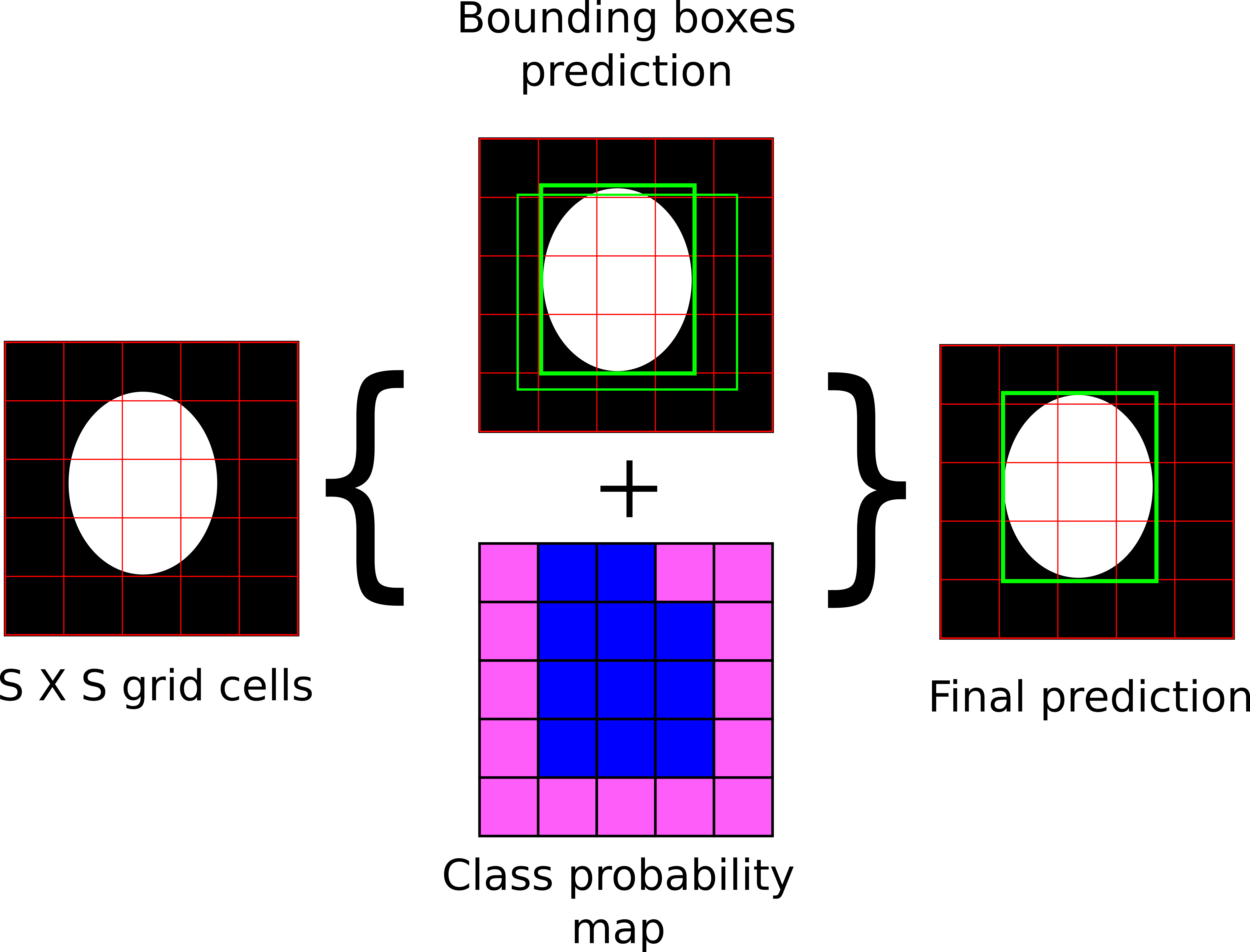}
\caption{\label{yolov3} Operating principle of the YOLO algorithm. Initially, an input image is split into $S \times S$ grid. In the next stage, multiple bounding boxes are predicted along with a class probability map (blue color highlights cells hosting an object belonging to the "droplet" class) to give final bounding boxes and classes of objects.  }
\end{figure}

The YOLO algorithm achieves high speed of image analysis due to its operating principle, namely by treating object detection as a regression instead of a classification task in determining class probabilities to the bounding boxes. The operating procedure of the YOLOv3 algorithm is sketched in Fig.~\ref{yolov3}. 
All objects in the given images are located and classified simultaneously. 
To achieve that, the YOLO algorithm divides the input image into $S \times S$ grid cells. 
Each grid cell is responsible for detecting an object if the object's center is within the cell domain. 
Each grid cell predicts $B$ bounding boxes with their confidence score for the object and $C$ conditional 
class probabilities for the given object belonging to a specific class. 
This information is then combined to produce the final output as a single bounding box around the 
detected object along with the class of that object. 
The final output is then passed to the DeepSORT algorithm for tracking the droplets. 

\subsection{DeepSORT}

DeepSORT is an algorithm that tracks detected objects between two successive frames~\cite{wojke}. 
By analyzing sequential frames, DeepSORT can construct trajectories of individual objects.  
It uses the Hungarian algorithm to distinguish the objects detected in 
two consecutive frames and assigns individual objects their unique identity. 
Kalman filtering is then used to predict the future position of the objects based on their current positions.

The YOLO and the DeepSORT algorithms together accomplish droplet recognition and tracking from the simulated systems. In the next section, we outline the training data acquisition process employed in this work and later describe the training process for the YOLO network for droplet detection.

\subsection{Training data}

Before training a network, acquiring training data is a crucial part of an object detection network. Training data consists of several images of the objects to be detected along with information about their location and dimensions. Practically, acquiring the training data is a labor-intensive task involving gathering the images and manually marking the positions and dimensions of the objects. For example, gathering thousands of images of street view cameras and marking the location of the cars to train a network to identify cars in a video feed. In addition, the prospect of manual marking of the objects makes the training data susceptible to human errors.

We avoided the labor-intensive process by preparing a synthetic training data set. In this dataset, each image is a collage with few solid white circles, mimicking the droplets placed randomly on a dark uniform background. The solid white circles of different sizes and shapes are prepared with commonly used computer graphics software. A separate text file is generated to note the positions and dimensions of the randomly placed droplets. 
A Python script generates several thousand images and associated text files in a matter of few seconds. The script is provided in the supplementary material [\textcolor{blue}{Link for the supplementary material}]. 
We prepared several types of training datasets, and the detailed description is given in Appendix~\ref{app:training_data}.

%The most apparent visual feature of the outcome of the L.B. simulations is the densely packed droplets of various sizes. 
We wish to emphasize that the training data, prepared synthetically, was intended to mimic the snapshots of the LB simulation, and it is not generated by a physical process associated with LB simulations. 
However, it is noteworthy that the synthetic data preparation method described above bypasses the labor-intensive 
data acquisition part and minimizes human errors, thus making the development process of the droplet tracking application much quicker and more precise. This method can potentially be used in several other object detection applications. 
In the next section, we describe the YOLO network's training process to identify the droplets with the synthetic data.

\subsection{Training}
\label{sec:training}
The training of the YOLO network is carried out through an iterative process. A subset of training data, called batch size, is passed through the network in each iteration, and then the network predicts the bounding boxes around the detected objects. 
A loss value is computed based on the difference between the output and the ground truth. 
The supplied label information is taken as the ground truth. The network parameters (weights and biases of the nodes) are updated to minimize the total loss and improve the accuracy of the network. For the YOLOv3 algorithm, the loss value, also called total loss, is the sum of three
separate contributions, regression loss, confidence loss, and classification loss~\cite{redmon}.

At regular intervals, i.e., after a few iterations, the accuracy of the network is assessed by computing mean average precision (mAP)~\cite{Everingham2010,henderson2017}. The mean average precision is a good measure to know how well the trained network performs on a dataset that is different from the training dataset. Typically, a separate dataset, called validation dataset, is compiled to compute the mean average precision.

We trained two YOLO networks to identify the droplets from the LB simulations. We adapted the code for training the network from Ref.~\cite{darknet}. The first network, called YOLO-tiny, is a lighter version of the full YOLO network consisting of fewer deep layers to trade training and inference speed for accuracy. For many practical purposes, the lighter version (YOLO-tiny) does perform reasonably well.

\begin{figure}[h]
\includegraphics[width= \textwidth, height=15cm,keepaspectratio]{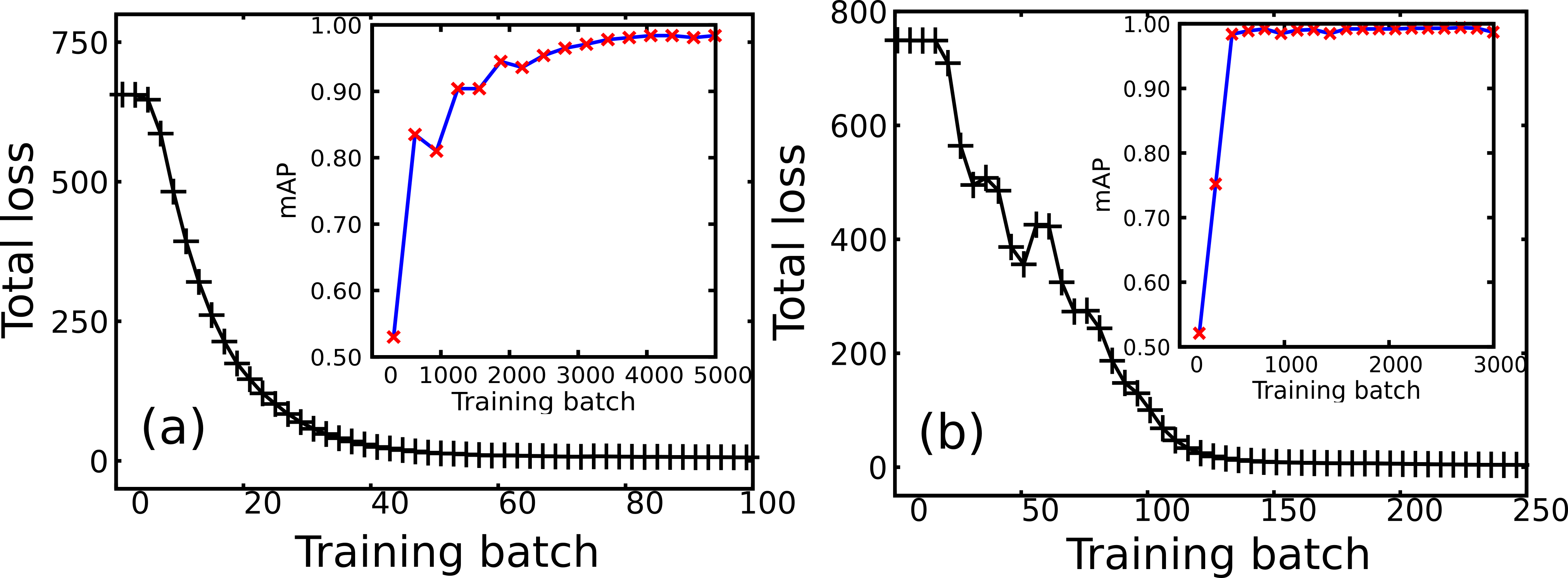}
\caption{\label{fig:map} Total loss and mean average precision (mAP) as the training progresses. (a) YOLO-tiny network, (b) YOLO network. }
\end{figure}

For training the YOLO-tiny network, we consolidated a total of $10000$ images, consisting of an equal number of two different types of synthetic images as shown in Fig.~\ref{fig:training}(a),(b). The parameters set for the training are mentioned here~\cite{para_list_tiny}.
The parameters of the network are updated at the end of every batch. A separate dataset of 1000 images, called validation data, was prepared from the same type of images to calculate the mean average precision. The total loss and mean average precision as the training progresses are shown in Fig.~\ref{fig:map}(a). 
As the training progresses, the total loss decreases, and the mean average precision of the network 
increases, indicating that the network is gaining performance with the training, reaching a total loss value very close to zero at the end of the training.
The mean average precision value is saturated to 0.985, indicating that the network is 98.5\% accurate in detecting the droplets in broad terms.

The second network we trained is the full YOLO network, with Darknet-53 as a backbone. 
Ten thousand images similar to the image shown in Fig.~\ref{fig:training}(c) were used as the training data. 
Another 1000 images of the same type were reserved as the validation data, and the training parameters are listed in ~\cite{para_list_full}. 
In this case as well, as the training progresses, the total loss decreases and mean average precision augments (as seen in Fig.~\ref{fig:map}(b)) saturating to 0.99, 
i.e. the network is 99\% accurate in detecting the droplets. 
The trained model in the form of a binary file containing the model's parameters (called weights) is saved. The weight file will be used later to analyze the data from LB simulations, and it is provided for 
both networks, the YOLO-tiny and the YOLO, as supplementary material [\textcolor{blue}{Link for the supplementary material}].

By training the YOLO network, we accomplish the first task of droplet recognition from the digital images, while the DeepSORT algorithm handles the second task of droplet tracking. In the next section, we analyze two LB simulations as case studies to test and
use the developed application.

\section{Results}
\label{sec:results}
After the model is trained with the desired accuracy in terms of mean average precision, it is ready to analyze the real-world data. From simulations or experiments, the real word images can be given as input to the trained network, and individual
droplets' trajectories can be obtained. This process is also called inference.

We run inference on the data generated with two LB simulations to test our trained network as described in Sec~\ref{sec:LB}. For inference, only the droplets (white color mass) were plotted on a dark background. The output video was then fed to the YOLO + DeepSORT network to infer the trajectories of individual droplets. We adapted code with Tensorflow implementation to run inference from Ref.~\cite{aiguy}.

\subsection{Multi-core emulsions}

\begin{figure}[!h]
\centering
\includegraphics[width= 10cm, height=15cm,keepaspectratio]{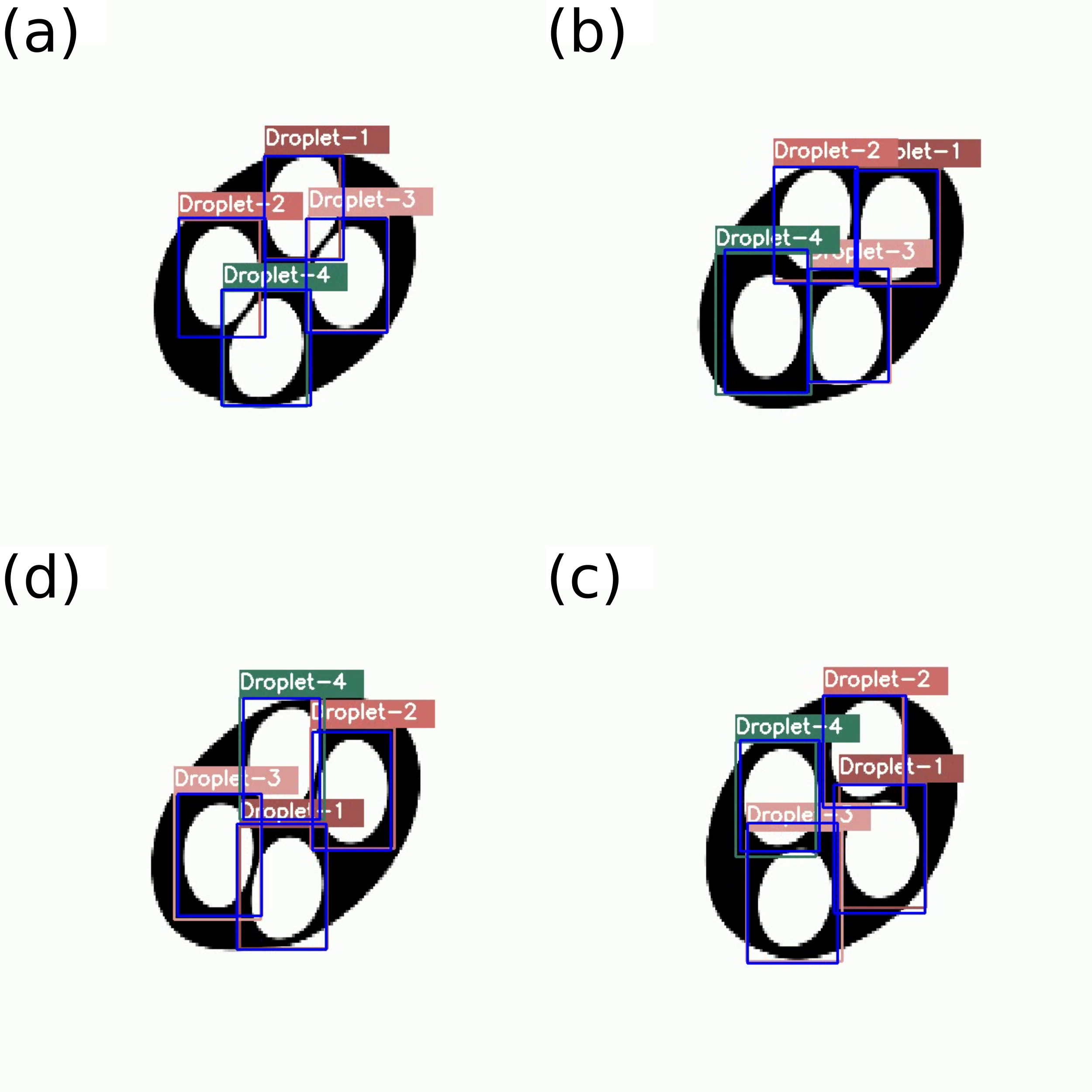}
\caption{\label{fig:at_detection1} Object detection and tracking at sequential frame number $N$. (a)$N=18$, (b)$N=19$, (c)$N=20$, (d)$N=21$. }
\end{figure}

Fig.~\ref{fig:at_detection1} shows a visual depiction of the object recognition by the YOLO network 
and tracking of the located objects by the DeepSORT network in four sequential images from the LB simulations. 
The bounding boxes predicted by the YOLO network are shown in blue color, and the unique ID assigned 
by the DeepSORT algorithm is written above each bounding box (see video3.avi).

\begin{figure}[!h]
\centering
\includegraphics[width= 13cm, height=15cm,keepaspectratio]{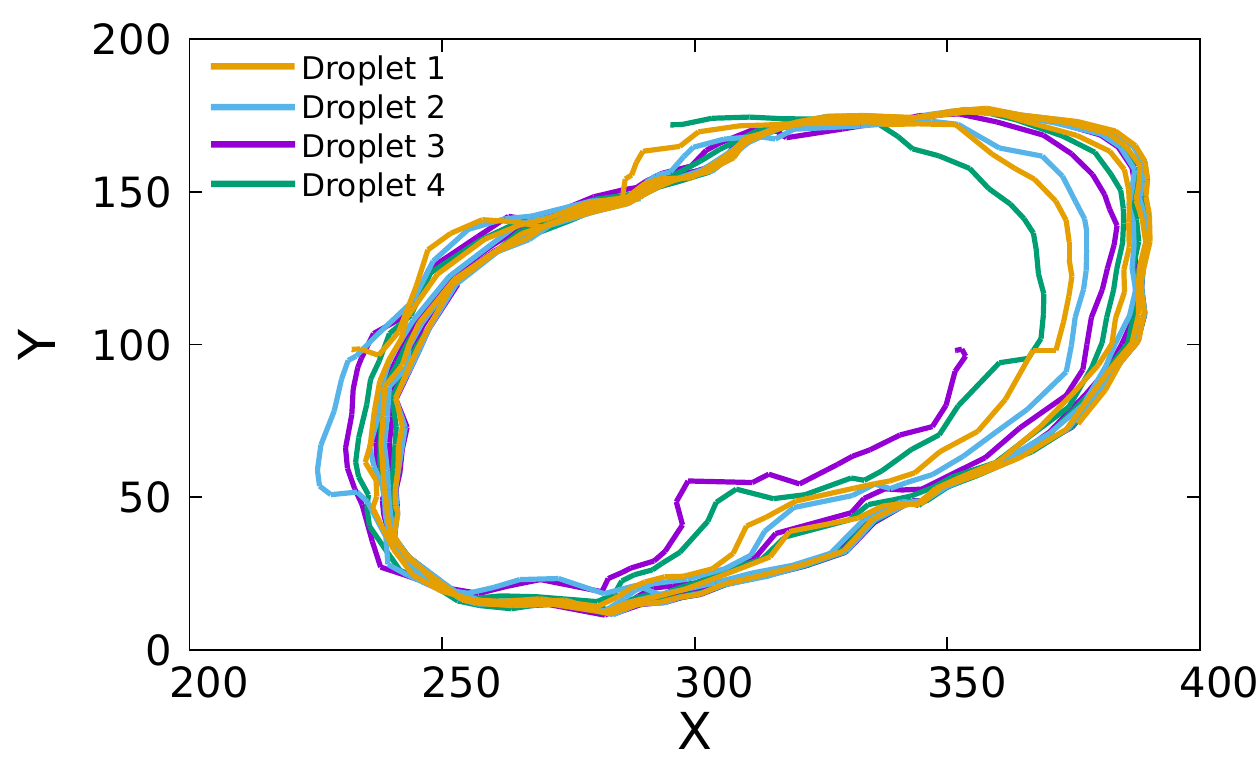}
\caption{\label{fig:at_4core_trajectory} Extracted trajectories of the center of mass of the individual droplets in a four core emulsion. On X-axis and Y-axis are the pixel numbers. }
\end{figure}

It is then straightforward to extract trajectories of the individual droplets across sequential frames since we have dimensions of all the bounding boxes in all the frames with their unique ID. The center of mass of the individual droplets is approximated as the geometric center of its bounding box. Fig.~\ref{fig:at_4core_trajectory} shows the paths traced by the droplets.

\subsection{Soft granular media}

In the second case, we run inference on LB simulation of soft granular media. 
This time, we employed the YOLO network to identify the droplets since the trained YOLO-tiny network makes many mistakes in droplet recognition. 
This is likely due to the increased complexity of the system, which presents a densely packed configuration of droplets moving through a narrow channel significantly deviating from its initial shape. 
 
\begin{figure}[!h]
\centering
\includegraphics[width= 13cm, keepaspectratio]{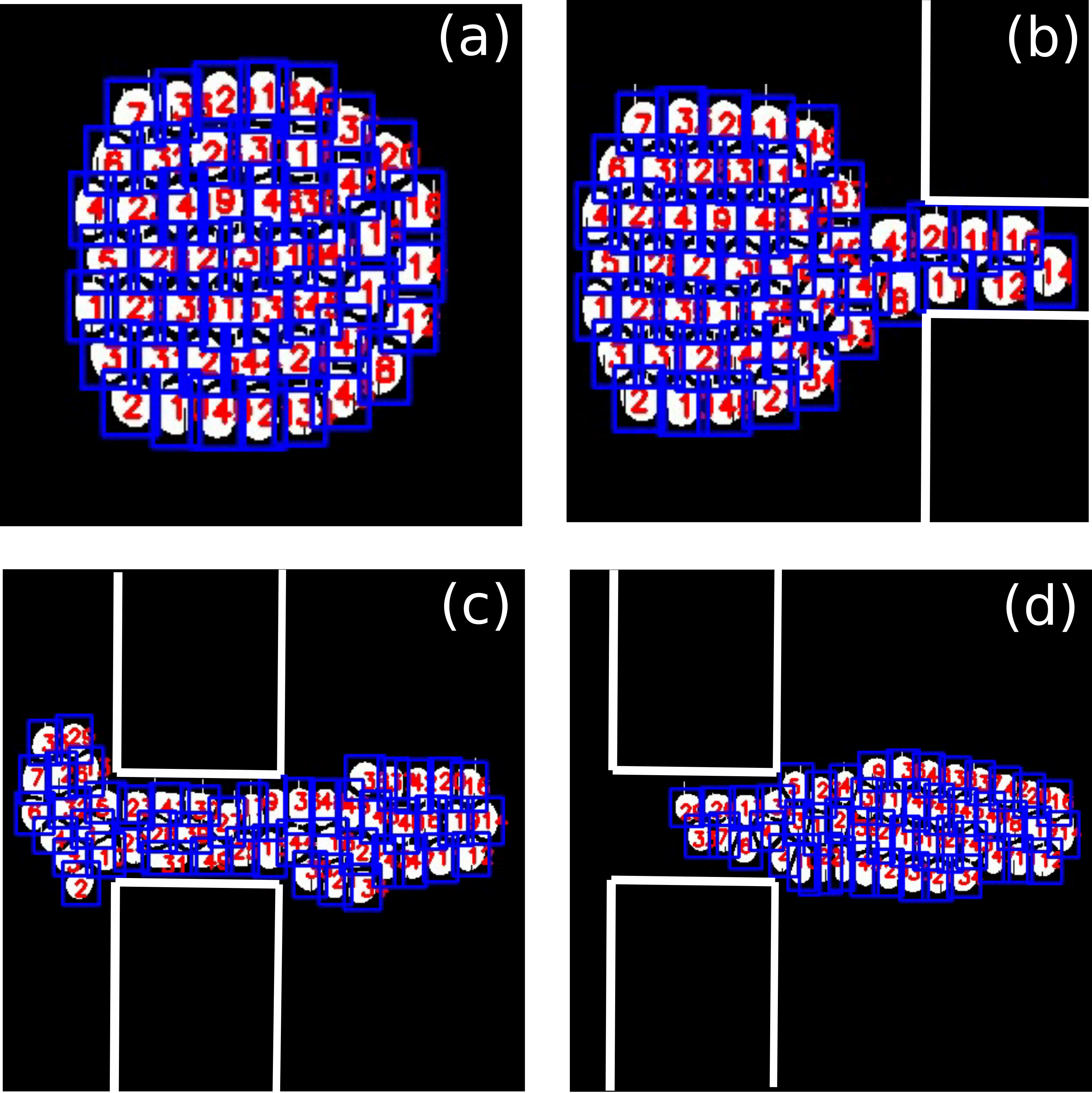}
\caption{\label{fig:am_video} Droplet recognition and tracking in four frames with frame number (a)$N=5$, (b)$N=6$, (c)$N=8$, (d)$N=9$. Blue bounding boxes are generated by the YOLO algorithm and unique IDs are assigned by the DeepSORT algorithm. }
\end{figure}

 The YOLO network, however, can locate the droplets with almost perfect accuracy. 
 Here, we note again that the training of the YOLO network was carried out with the training data consisting of randomly placed dense clusters of solid white circles. As before, Fig.~\ref{fig:am_video} shows the droplet identification by the YOLO network and tracking with the DeepSORT algorithm. 
 The recognized droplets are shown with blue bounding boxes, and the unique IDs assigned by the DeepSORT algorithm are placed within the bounding box for visual inspection (see video4.avi). 

\begin{figure}[!h]
\centering
\includegraphics[width= 13cm, height=15cm,keepaspectratio]{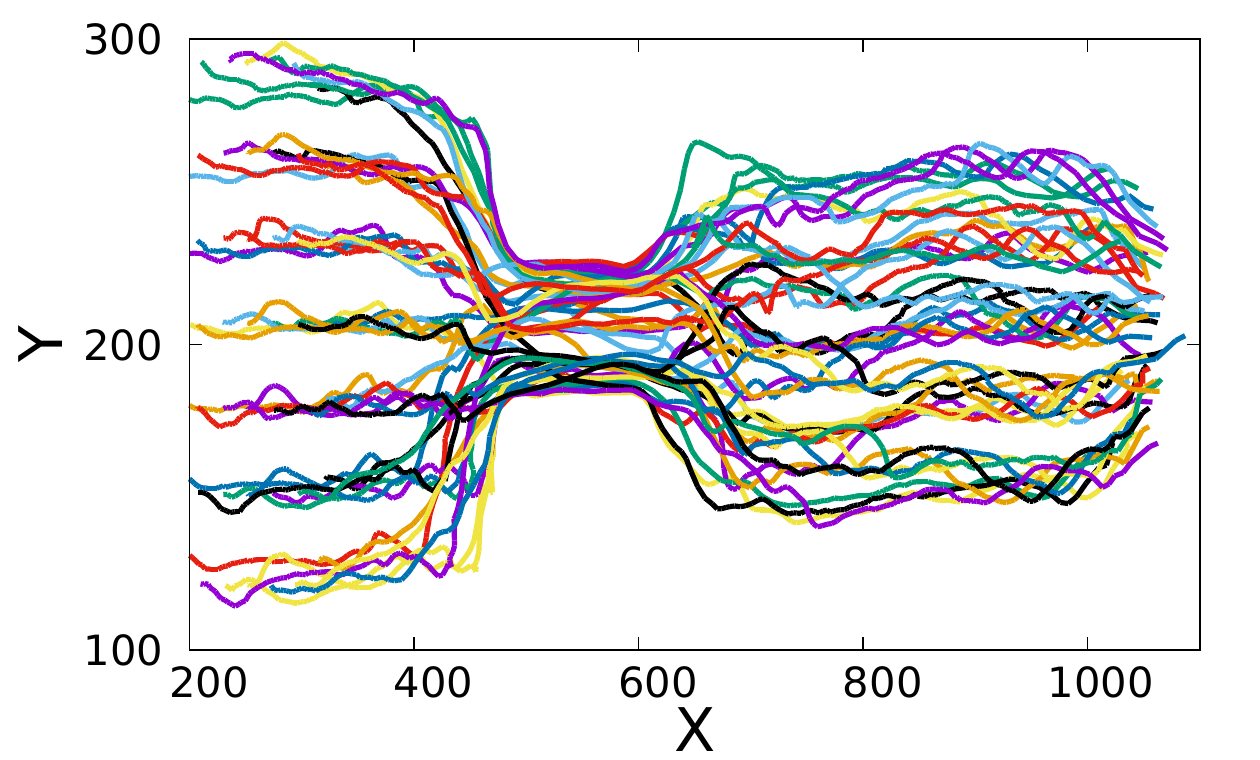}
\caption{\label{fig:am_trajectory} Extracted trajectories of the center of mass of the individual droplets represented by unique colors. X and Y axis show the pixel number.}
\end{figure}

In Fig.~\ref{fig:am_trajectory}, we plot the extracted trajectories of the center of mass of the individual droplets. The center of the bounding box is approximated as the center of mass. In this case, we obtain 49 trajectories for the 49 droplets. Each separate trajectory is shown with a unique color.

\subsection{Accuracy of the trained model}

During the training, validation data were used to estimate the accuracy of the object detector model. 
In our case, they were statistically identical to the training dataset and could provide a biased estimate of the accuracy. 
Unbiased accuracy of the trained model can be obtained by analyzing a test case (also called test data), ideally from 
a real-world dataset with known droplet trajectories. 
From the output trajectories of the model and the
known trajectories as ground truth, the accuracy of the trained model can be estimated for real-world applications.

To measure the accuracy of the trained model, we compute mean square errors (MSE) by a Euclidean metric: 
\begin{equation}
  MSE  =  \frac{1}{N}\sum_{i=1}^{N} (y_{cm}(i) - \hat y_{cm}(i))^2.
\end{equation}

Here, $N$ is the total number of frames, $y_{cm}$ is the y coordinate of the center of mass computed by some other independent way, $\hat y_{cm}$ is the inferred y coordinate of the center of mass by the deep learning tool.

\begin{figure}[!h]
\centering
\includegraphics[width=\textwidth, height=15cm,keepaspectratio]{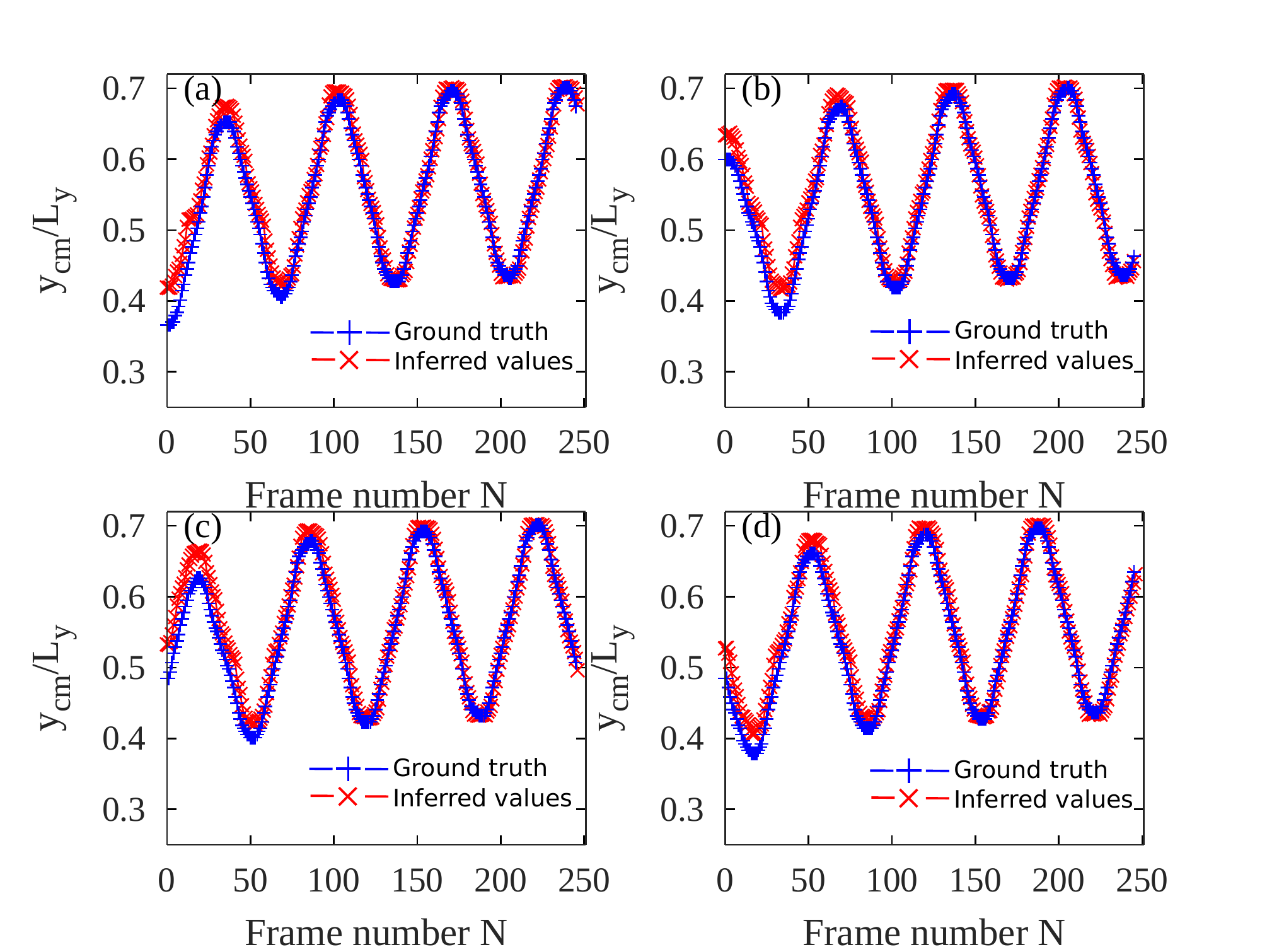}
\caption{\label{fig:at_ground_truth} (a)-(d) Comparison of the inferred center of mass of the four individual droplets in every frame with the ground truth computed by the analytical expression in Ref.~\cite{tiribocchi_prf}. Mean square error $MSE$ between the inferred values and the ground truth values is computed to measure the error. (a) $MSE = 3.44 \times 10^{-2}$, (b) $MSE = 1.80 \times 10^{-2}$, (c) $MSE = 3.12 \times 10^{-2}$, (d) $MSE = 1.59 \times 10^{-2}$.  }
\end{figure}

We measure the accuracy of the YOLO-tiny + DeepSORT model by comparing the inferred trajectories of the droplets with the known ground truth.
In this case, the ground truth is the center of mass computed with an analytical expression (see Ref.~\cite{tiribocchi_prf}).
In Fig.~\ref{fig:at_ground_truth}, we compare the ground truth  $y$-coordinates of the center of mass ($y_{cm}$) of individual droplets with the inferred $y$-coordinate 
values ($\hat y_{cm}$) by the deep learning tool. 
We observe that the deep learning model agrees with the analytical expression after a few initial frames.
However, in the initial frames, the deep learning model slightly deviates from the true values, even though both the curves are in qualitative agreement.
The mean square value $MSE$ is of the order of $10^{-2}$ for all four tracked droplets, indicating a good agreement between the ground truth and the inferred values.

\begin{figure}[!h]
\centering
\includegraphics[width=\textwidth, height=15cm,keepaspectratio]{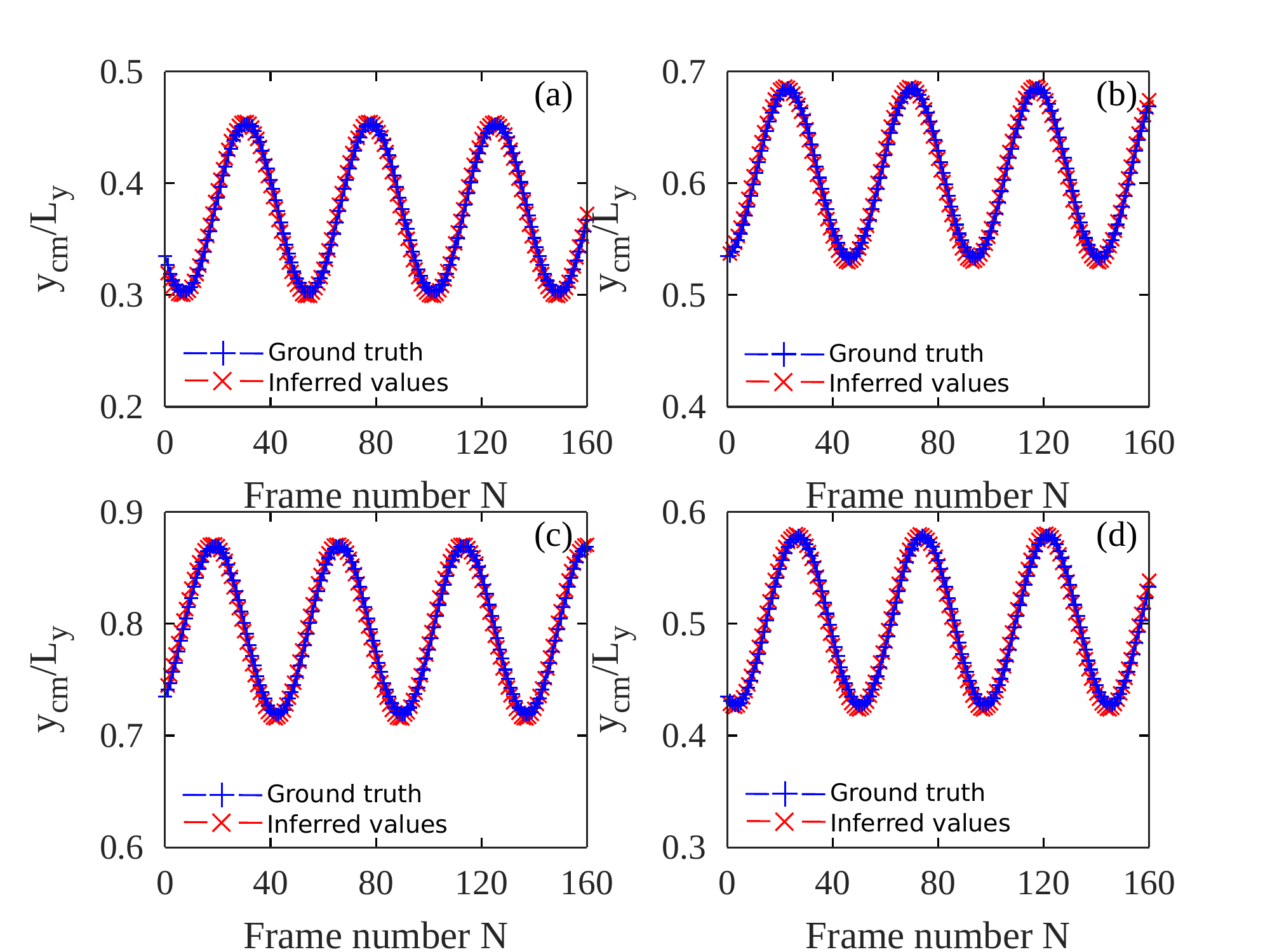}
\caption{\label{fig:am_ground_truth} Measuring accuracy of the YOLO + DeepSORT model. (a)-(d) Inferred path of four different droplets compared with their predefined paths. Mean square error $MSE$ between the inferred values and the ground truth values is computed to measure the error. (a) $MSE = 1.41 \times 10^{-4}$,(b) $MSE = 1.52 \times 10^{-4} $, (c) $MSE = 1.55 \times 10^{-4} $, (d) $MSE = 1.45 \times 10^{-4} $.  }
\end{figure}

For measuring the accuracy of the full YOLO + DeepSORT model, we prepared a synthetic test case where the center of mass of the droplets move along some predefined paths (see Appendix~\ref{app:sythetic_tes_data}). The predefined paths serve as the ground truth.  We employed the YOLO + DeepSORT model to infer trajectories of the droplets from the video and the comparison is shown in Fig.~\ref{fig:am_ground_truth}. Four droplets were tracked and the inferred path is compared with the known predefined trajectories of these droplets. Both the trajectories are in excellent agreement and we see no deviation in the inferred and true trajectories in any phase of the tracking. The YOLO + DeepSORT model is highly accurate with MSE of the order of $10^{-4}$. The low values of mean square errors once again point to the high accuracy of the trained model. 

We emphasize that we measure the accuracy of the developed application, including the object recognition and the object tracking process. The error values are indicative of the total expected error in extracted trajectories from the real-world data. Apart from the model's accuracy, the model's analysis speed is an important factor to consider while developing an application for practical use. In the next section, we report the inference speed, i.e., the rate of image analysis by the developed application on different computer hardware configurations.

\subsection{Inference speed}

%Inference speed, i.e., the time required for the algorithm to process the input and produce the output, is an important factor from the practical deployment of the developed application. \textcolor{red}{AT: I would remove this sentence and start with the next one.}

We test the inference speed of the object detection and tracking model on a 
typical notebook computer and a commonly used GPU machine. 
For inference, we used Tensorflow implementation~\cite{aiguy}. The inference speed, analyzed frame per second (FPS), for the YOLO-tiny and the YOLO network are tabulated in table~\ref{table1}.

\begin{table}[!h]
    \centering
    \begin{tabular}{|| c | c | c ||}
       \hline
       %\multirow{\textbf{Machine}}
       \multicolumn{1}{||c|}{\textbf{Network/Machine}}
       &\multicolumn{2}{c||}{\textbf{Inference speed}}\\
       \cline{2-3}
    & \textbf{CPU (i3-2328M)} & \textbf{GPU (RTX 2060 Super)} \\
 \hline\hline
        YOLO-tiny + DeepSORT &  2.44 fps & 35 fps \\
        YOLO + DeepSORT & 0.25 fps & 12 fps \\
        \hline
    \end{tabular}
    \caption{Average inference speed in frames per second (fps) on two hardware configurations. }
    \label{table1}
\end{table}

These data clearly show that there is room for improvement and optimization of the inference speed for both YOLO and DeepSORT, for instance, by coding them in other programming languages, such as C++. This is an interesting topic for future work.

Having accurate trajectories of moving objects is the first step to study the system. In the next section, we present an analysis of the inferred trajectories to test a hypothesis.   

\subsection{Flocks of droplets?}

%Obtaining the trajectories of the droplets is the first step towards further analysis. \textcolor{red}{AT: I would also remove this first sentence and start from the next one, unless we specify a bit more on the analysis.}

At first sight, the droplets seem to move like the agents in active matter systems, such as birds in a flock. 
Several statistical physics models have been proposed to explain the flocking behavior in animal groups~\cite{vicsek1995, couzin2002, Cavagna2015}, and experimental studies involving observation of moving animal groups inferred the interaction rules between 
the agents~\cite{Herbert-Read18726,Lukeman12576}. One prominent feature common in most of these models is that 
the agents (birds) move along the average direction of their neighbors. 

\begin{figure}[!h]
\centering
\includegraphics[width= \textwidth, keepaspectratio]{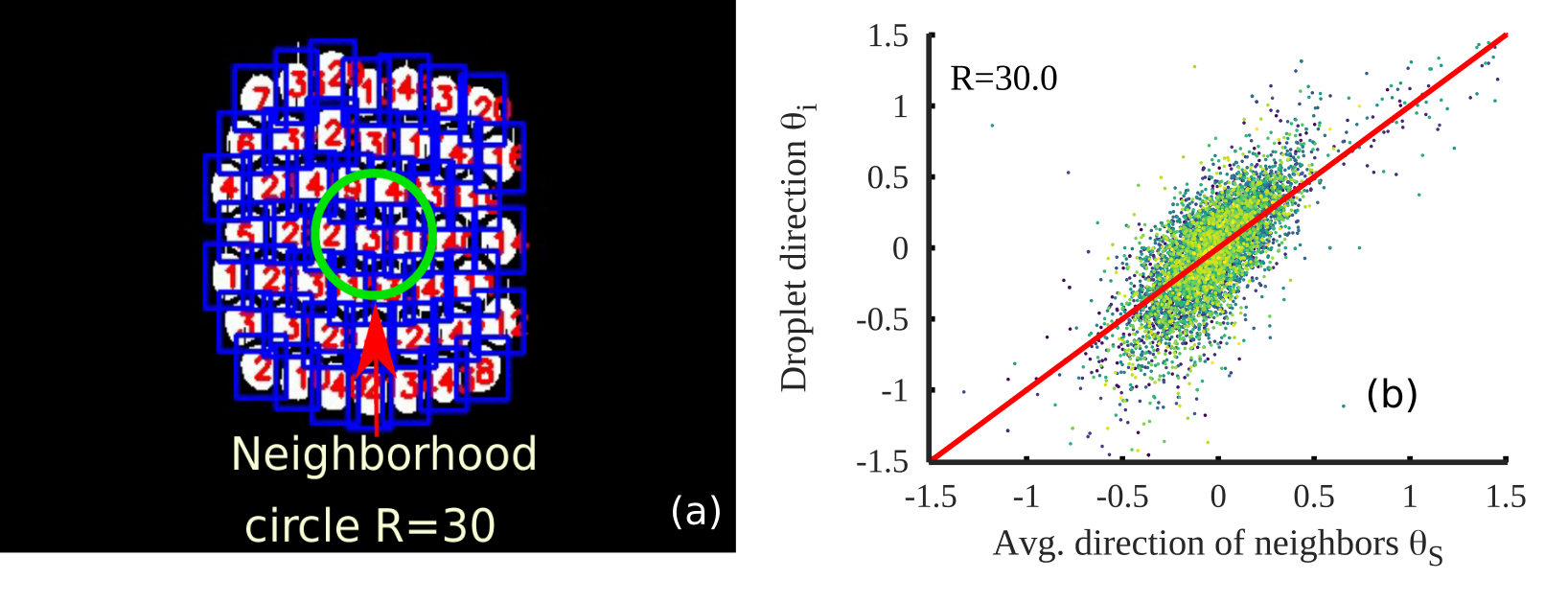}
\caption{\label{fig:analysis1} (a) Neighborhood of one of the randomly chosen droplet. The neighborhood of a droplet is a circle with $R=30$ pixels. (b) Scatter plot of the direction of individual droplets $\theta_i$ with the average moving direction of droplets $\theta_S$ in the neighbourhood of the droplet $i$. Values on the axes are in radians. The thick red line is a guideline to identify instances when the moving direction of droplets and their neighbors are identical.}
\end{figure}

Given the superficial similarity between a flock and the soft granular media system studied here, 
we perform a quick analysis to test a hypothesis that the droplets behave like agents in an active matter system. 
To this end, we calculate the average moving direction $\theta_S$ of the neighbors of 
each individual droplet. The neighbors of droplet $i$ are defined as droplets within a circle of radius $R$, fixed on the 
center of the droplet $i$. We set the size of the neighborhood in such a way that few neighboring droplets are included  (see Fig.~\ref{fig:analysis1}(a)). Fig.~\ref{fig:analysis1}(b) shows a scatter plot between individual droplet's moving direction $\theta_i$ and the average direction of their neighbors $\theta_S$ for the entire duration of the simulation. The moving direction is measured as angle $\theta$ made by the velocity vector with the x-axis in a fixed frame of reference. Thus, the range of $\theta$ is from $-\pi$ to $+\pi$. Each different color in this plot shows a separate identity of the individual droplets. The red line is a guideline for the eyes to show the situation when the individual droplet's moving direction is identical to that of its neighbors. 

In Fig.~\ref{fig:analysis1}, around 47\% of all the total points are in the band when $\vert \theta_i - \theta_S \vert < 0.12$. i.e., when the angle between the droplet's motion and its neighbors is less than $7$ degrees. This suggests that the moving direction of individual droplet deviates 
(by more than $7$ degrees) from the average direction of their neighbors 53\% of the time. 
We conclude that the system of droplets does not move like self-propelled particles in a quantitative sense. 
This is plausible since, besides some similarities, there are also several points of departure between
the two systems.  
First,  in soft granular media systems, the fluid flow carries the droplets along the flow direction, like a systematic wind and 
second, the wall boundaries impose a stringent constraint  on the motion of the droplets. 
None of the two aspects above are generally considered in the modeling of active matter systems. 
It thus appears that the motion of the droplets cannot be quantitatively cast in terms of simple rules which hold for flocks \cite{Giardina2008}.
This raises an interesting question on how to generalize the flock equations to come up with a set of effective
equations of the droplets in complex many-body flows such as the one explored here. 
We plan to address this question in future work.

\begin{comment}

\subsection{Inference speed}

Inference speed, i.e., the time required for the algorithm 
to process the input and produce the output, is an important factor from the practical deployment of the developed application. 

We test the inference speed of the object detection and tracking model on a 
typical notebook computer and a commonly used GPU machine. 
For inference, we used Tensorflow implementation~\cite{aiguy}. The inference speed, analyzed frame per second (FPS), for the YOLO-tiny and the YOLO network are tabulated in table~\ref{table1}.

\begin{table}[!h]
    \centering
    \begin{tabular}{|| c | c | c ||}
       \hline
       %\multirow{\textbf{Machine}}
       \multicolumn{1}{||c|}{\textbf{Network/Machine}}
       &\multicolumn{2}{c||}{\textbf{Inference speed}}\\
       \cline{2-3}
    & \textbf{CPU (i3-2328M)} & \textbf{GPU (RTX 2060 Super)} \\
 \hline\hline
        YOLO-tiny + DeepSORT &  2.44 fps & 35 fps \\
        YOLO + DeepSORT & 0.25 fps & 12 fps \\
        \hline
    \end{tabular}
    \caption{Average inference speed in frames per second (fps) on two hardware configurations. }
    \label{table1}
\end{table}

These data clearly show that there is room for improvement and optimisation of the inference speed
for both YOLO and DeepSORT, for instance by coding them in other programming languages, such as C++.
This is also an interesting topic for future work. 

\end{comment}

\section{Conclusions}
\label{sec:conclusions}

We adapted and combined deep learning-based object recognition algorithm YOLO 
and object tracking algorithm DeepSORT to analyze data generated by fluid simulations of two 
soft flowing systems, multicore emulsions, and soft granular media. 
In particular, the trajectories of the individual droplets moving within a microchannel flow 
were extracted from the digital images of the system. 

We determine the accuracy of the complete application by comparing the inferred 
trajectories with the ground truth trajectories computed by independent methods. 
Although we used LB simulation data as a case study; the developed application  
can also be applied data generated by two-dimensional experimental setups. 

With the use of commonly available GPUs, the YOLO + DeepSORT based application can analyze 
the images at a remarkable rate, above the image capture rate of typical cameras, thus offering  a practical, low-cost application capable of analyzing real-time data as it is acquired. 
 
 From the technical point of view, we showed that the synthetically prepared datasets could be used to train the object detection network with almost perfect accuracy, thus avoiding labor-intensive, time-consuming 
 training data acquisition for object recognition network training. 
 %Also, a synthetically prepared unbiased test case can be used to measure the accuracy of the complete application.

It is hoped and expected that further developments of the work presented here may pave the way
to the automatic detection and tracking of moving agents in complex 
flows of scientific, engineering, and biological interest.

\section{Acknowledgement}
M. D., F. B., A. M., M. L., A. T. and S. S. acknowledge funding from the European Research Council under the European Union's Horizon 2020 Framework Programme (No. FP/2014-2020) ERC Grant Agreement No.739964 (COPMAT).

\appendix

\section{Color-gradient approach with near contact interactions}
\label{app:color_grad}

Here we describe the LB approach used to simulate the translocation through a constriction of a high internal phase double emulsion. We employ a color-gradient \cite{leclaire2012numerical,leclaire2017,gunstensen1991,montessori2018regularized,montessoriprf,rothman1988immiscible} regularized LB method \cite{montessori2015lattice,latt2006lattice,coreixas2019comprehensive}, which is built starting from  two sets of probability distribution functions capturing the dynamics of the immiscible fluid components. Each set evolves via a sequence of streaming and collision steps \cite{kruger2017lattice,succi2018lattice,succi2015lattice,benzi1992lattice}
\begin{equation} \label{LBM1}
f_i^k(\mathbf{x}+\mathbf{c}_i \Delta t, t+\Delta t)=f_i^k(\mathbf{x},t) + \Omega_i^k (f_i^k(\mathbf{x},t)) + F_i^{rep},
\end{equation}
where $f_i^k(\mathbf{x},t)$ is the $i^{th}$ discrete probability distribution function for the $k^{th}$ component, giving the probability of finding a fluid particle at position $\mathbf{x}$, time $t$ and with discrete velocity $\mathbf{c}_i$. The index $k$ is such that $k=1,2$, while $i$ belongs to the range $0 \le i \le N_{set}$, where $N_{set}$ is the dimension of the set of discrete probability distribution functions and is equal to $8$ for the two-dimensional nine speeds lattice (D2Q9) employed in this paper.
The time step $\Delta t$ is expressed in lattice units \cite{kruger2017lattice} and set to $1$, a usual choice in the LBM \cite{succi2018lattice}. Finally, $F_i^{rep}$ is a force aimed at upscaling  the repulsive near-contact forces acting on scales much smaller than the resolved ones \cite{montessori2019mesoscale,montessori2019modeling}.

As stated above, the multicomponent approach employed in this work is based on a variant of the color-gradient LB model. Following the standard formalism, the  collision operator can be split into three parts
\begin{equation} \label{LBM4}
\Omega_i^k = (\Omega_i^{k})^{(3)} [(\Omega_i^{k})^{(1)}+(\Omega_i^{k})^{(2)}].
\end{equation}
The first term $(\Omega_i^{k})^{(1)}$ is the usual single relaxation time Bhatnagar–Gross–Krook collisional operator \cite{succi2018lattice} while the second part $(\Omega_i^{k})^{(2)}$ is the \textit{perturbation step}, which accounts for the interfacial tension, and reads as
\begin{equation} \label{LBM8}
(\Omega_i^{k})^{(2)}= \frac{A_k}{2} |\nabla{\Theta}|\Bigg( w_i \bigg( \frac{\mathbf{c}_i \cdot \nabla{\Theta}}{|\nabla{\Theta}|} \bigg)^2-B_i \Bigg),
\end{equation}
where $A_k$ ($k=1,2)$ and $B_i$  $(i=0,N_{set})$ are suitable constants defined in \cite{montessori2019mesoscale}. In Eq.(\ref{LBM8}), $\Theta$ is a scalar phase field, defined as 
\begin{equation} \label{LBM9}
\Theta=\frac{\rho_1-\rho_2}{\rho_1+\rho_2} 
\end{equation}
assuming the value $1$ in the fluid component with density $\rho_1$ and the value $-1$ in the fluid component with density $\rho_2$. It is worth observing that the constants $A_1, A_2$ are related to the surface tension $\sigma$ by
\begin{equation} \label{LBM10}
\sigma=\frac{2}{9} \tau (A_1+A_2).
\end{equation}
The third term $(\Omega_i^{k})^{(3)}$ is the recolouring operator, as described in \cite{latva2005diffusion},  which aims at minimizing the mutual diffusion between the fluid components, thus favouring their separation.

Finally, the last term at the right hand side of Eq.(\ref{LBM1}) codes for the short-range, repulsive force at the interface between the two fluid components, 
%localized on the interface 
aimed at frustrating the coalescence between interacting, neighboring droplets.

\section{Free energy LB models for interacting multicomponent fluids}
\label{app:free_energy}

In this subsection we shortly outline the multicomponent field model for immiscible fluid mixtures employed to simulate the physics of three and four-core emulsions. Further details can be found in \cite{marenduzzo_prl,marenduzzo_soft,tiribocchi_pof,tiribocchi_prf,tiribocchi_nat}.

In this approach, a set of scalar phase-fields $\psi_i({\bf r},t)$, $i=1,....,M$ (where $M$ is the number of cores) is used to model the droplet density (positive within each drop and zero outside), while the average fluid velocity of both drops and solvent is described by a vector field ${\bf v}({\bf r},t)$. 
 
The dynamics of the fields $\psi_i({\bf r},t)$ is governed by a set of convection-diffusion equations
\begin{equation}\label{CH_eqn}
\partial_t\psi_i+{\bf v}\cdot\nabla\psi_i=D\nabla^2\mu_i,
\end{equation}
where $D$ is the mobility, $\mu_i=\frac{\delta{\cal F}}{\delta\psi_i}$ is the chemical potential and ${\cal F}=\int_V fdV$ is the total free energy describing the equilibrium properties of the fluid suspension.
The free energy density $f$ is given by \cite{degroot,lebon}
\begin{equation}\label{freeE}
f= \frac{a}{4}\sum_i^M\psi_i^2(\psi_i-\psi_0)^2+\frac{k}{2}\sum_i^M(\nabla\psi_i)^2+\epsilon\sum_{i,j,i<j}\psi_i\psi_j,
\end{equation}
where the first term is the double-well potential ensuring the existence of two coexisting minima, $\psi_i=\psi_0$ inside the $i$th droplet and $\psi_i=0$ outside, and the second term stabilizes the droplet interface. The parameters $a$ and $k$ are two positive constants related to the surface
tension $\sigma=\sqrt{8ak/9}$ and the width of interface 
 $\xi=2\sqrt{2k/a}$ \cite{widom,kruger2017lattice}.
Finally, the last term in Eq.(\ref{freeE}) is a soft-core repulsion contribution which penalizes the overlap of droplets, and whose strength
is tuned by the positive constant $\epsilon$.

The fluid velocity ${\bf v}$ obeys the Navier-Stokes equation, which in the incompressible limit reads
\begin{equation}\label{nav_stok}
\rho\left(\frac{\partial}{\partial t}+{\bf v}\cdot \nabla\right){\bf v}=-\nabla p+\eta\nabla^2{\bf v}-\sum_i\psi_i\nabla\mu_i.
\end{equation}
Here $\rho$ is the fluid density, $p$ is the isotropic pressure and $\eta$ is the dynamic viscosity. 
A typical set of values of thermodynamic parameters used in simulations is the following: $a=0.07$, $k=0.1$, $D=0.1$, $\eta=1.67$, $\epsilon=0.05$, $\Delta x=1$ (lattice step), $\Delta t=1$ (time step).

Equations (\ref{CH_eqn}) are solved using a finite difference scheme while Eq. (\ref{nav_stok}) is integrated by means of a standard lattice Boltzmann method. The latter shares many features with the LB model described in the previous section. Indeed, it is built from a set of distribution functions whose evolution is governed by a discrete Boltzmann equation akin to Eq.\ref{LBM1} and defined on a D2Q9 lattice. In both methods it can be shown that, once conservation of mass and momentum are fulfilled, the Navier-Stokes equation can be obtained by performing a Chapman-Enskog expansion of the distribution functions \cite{succi2018lattice,tiribocchi_epje}. However, unlike the previous algorithm, in this approach the thermodynamics is encoded in a free-energy employed to calculate the forces (chemical potential and pressure tensor) controlling the relaxation dynamics of the mixture. In addition, this model allows for an immediate tracking of position and speed of the droplets by computing the coordinates of the center of mass.

\section{Training data}
\label{app:training_data}
Training data is used to train the YOLO network to locate the droplets in given input images. Now we describe the training data acquisition employed in this work. Training data for the YOLO network must consist of several images of the objects of interest and their positions in the images. Training data is passed through the network, and the network parameters are updated with every pass to increase the detection accuracy. In a typical exercise to train a network, training data is compiled by acquiring images taken by digital cameras in the real world. Objects in those images are manually identified and marked with bounding boxes encapsulating each object. A separate label file containing the class of the object, positions (x,y coordinates), and dimensions (width and height) of the bounding boxes is prepared. %Preparing such a training dataset is highly labor-intensive, and it is susceptible to errors in marking the objects.
%For example, if YOLO network is to be trained to identify cars from the street-view images then typically few thousand images containing various makes and models of cars are collected. In each image, location of each car is marked by means of a bounding boxes with each bounding box encapsulating a single car. In associated text file, called the label of the image, the dimension of each bounding box such as the width w, height h, x-coordinate x, and y-coordinate y are noted down along with the class number of the object it encapsulates. Several such images with their associated text file containing the labels form a training dataset for the YOLO network.

%In order to train a YOLO object recognition network, a training data need to be compiled in a specific format. Typically, multiple images containing the objects of interest are collected. 

We avoided the labor-intensive part of image acquisition and manual labeling by preparing a synthetic training dataset. The training dataset must contain several images that bear some visual features of the expected outcome of the LB simulations, as shown in Fig.~\ref{fig:translocation} and Fig.~\ref{fig:emulsion}. In our case, the outcome of the LB simulations are the images with the area occupied by the droplets shown by white color on a high contrast dark background. We prepared our training dataset by randomly placing solid white circles on a dark uniform background. The solid white circles were intended to resemble the droplets, and the high contrast background is intended for easy identification of the droplets from the background. 

%In this work, we analyze two L.B. simulations. \textcolor{blue}{Some small description here}
\begin{figure}[h]
\includegraphics[width= \textwidth, height=15cm,keepaspectratio]{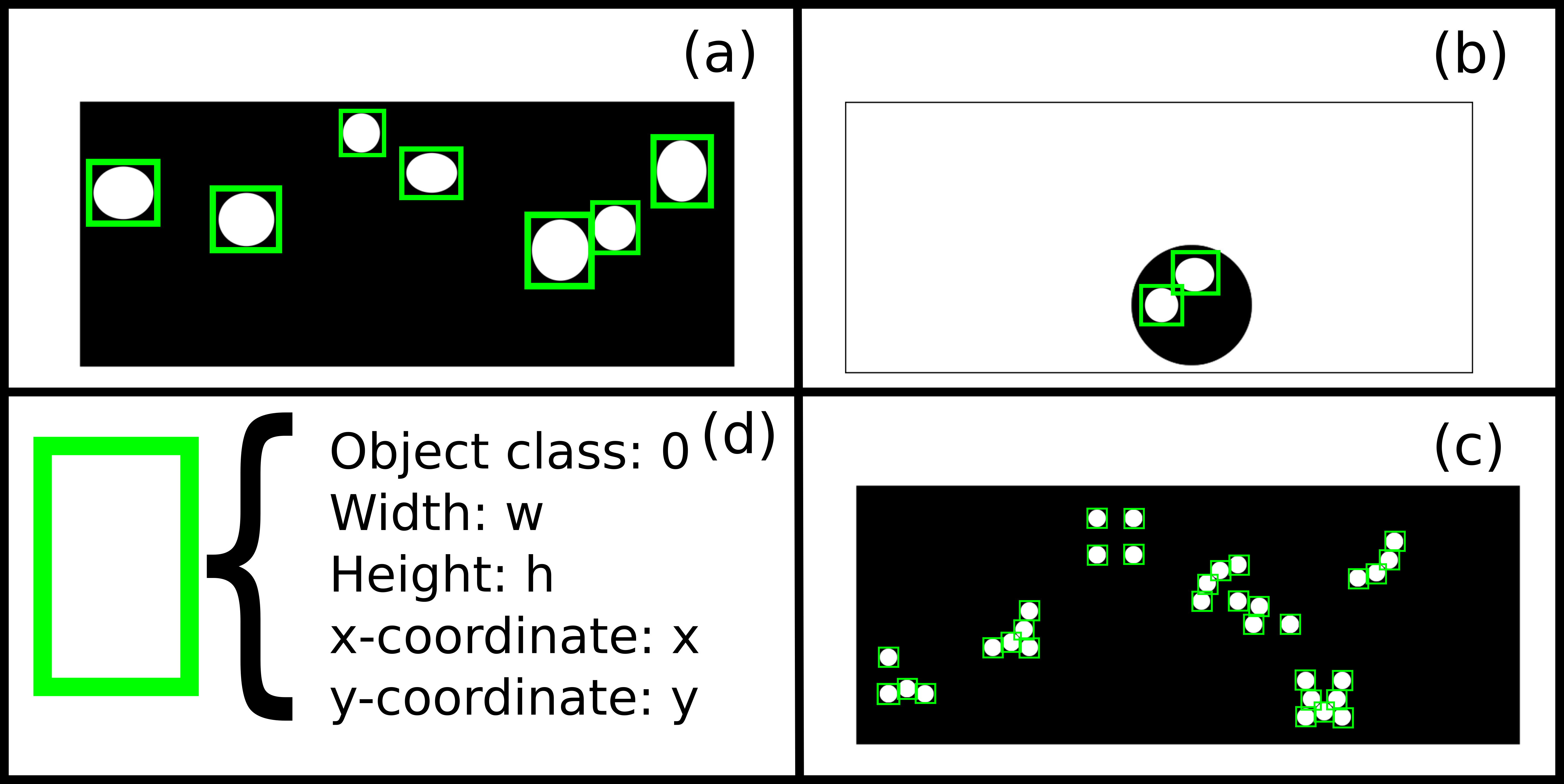}
\caption{\label{fig:training} (a) Randomly placed droplets on a black uniform background of size $1200 \times 600$. (b) Two droplets are randomly placed in a black envelope. The placement of the envelope is random within the white uniform background of size $1200 \times 600$ pixels. (c) Few random size clusters of droplets were placed randomly in a uniform background of size $1200 \times 600$. (d) Label information associated with each of the bounding boxes.}
\end{figure}

In the case of LB simulation of multi-core emulsions, the outcome consists of few elliptical shape droplets enclosed in a dark bag on a uniform white background. We incorporate these features in the training data set by preparing 10000 images each of two types of images. The first type of images (see a representative Fig.~\ref{fig:training}(a)) are prepared by placing the white solid circles and ellipses of randomly chosen semi-minor and semi-major axis on the background of size 1200 $\times$ 600 to teach the network to identify droplets of various sizes. In the second type of images (see Fig.~\ref{fig:training}(b)), two solid circles as droplets are placed randomly within a solid black circle. The position of the solid black circle is chosen randomly in the uniform white background. %These images are similar in appearance from the one we obtain via L.B. simulations of the multi-core emulsions. 

In the case of LB simulations of soft granular media, the most apparent feature is the dense droplets cluster. In this case,  few clusters with the random number of solid white circles are placed in a uniform solid black background to compile training data. The training data consisted of 10000 such images (see Fig.~\ref{fig:training}(c) as an example).

The label information for every image is shown in Fig.~\ref{fig:training}(d). Each bounding box encloses a single white droplet. In associated text files, the dimensions of the bounding boxes (x and y coordinates, width and height, and the class of object it encapsulates) are noted down. Images and the training data together form a training dataset for the YOLO network.
%We emphasize here that the training dataset does not contain the actual images from the L.B. simulations. The training dataset contains the synthetically prepared images to train the network to identify the droplets from real L.B. simulations.

\section{Synthetic test case}
\label{app:sythetic_tes_data}

To measure the accuracy of the developed YOLO + DeepSORT application, we prepared
a test video. In this video, the center of mass of each droplet move along predefined trajectories (see video5.avi). The following equation gives the trajectory of individual droplets;

%\begin{equation}
%y^i =B \hspace{1mm} sin(kx)
%\end{equation}

\begin{equation}
\label{eq:am_vid}
\begin{split}
x^i(t+1) = x^i(t) + A, \\
y^i(t+1) = y^i(t) + B \hspace{1mm} sin(kx^i(t)).
\end{split}
\end{equation}

Here, $x^i(t)$ and $y^i(t)$ are the coordinates of the $i^\text{th}$ droplet at time t. $A$, $B$ and $k$ are the suitable constants set as $A=5$, $B=5$ and $k=5$. 

\begin{figure}[h]
\includegraphics[width= 12cm, height=15cm,keepaspectratio]{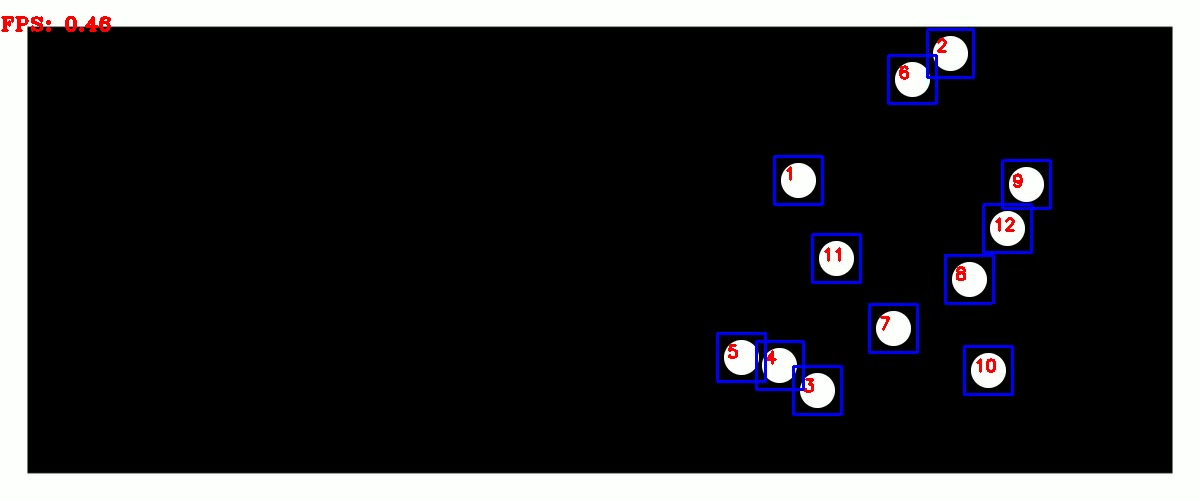}
\caption{\label{fig:sythetic_track} Snapshot of the output from YOLO + DeepSORT network to track droplets moving on predefined paths. Blue boxes are predictions by the YOLO network, and the DeepSORT network assigned unique ids.}
\end{figure}

In practice, we prepared several images and in each image we pasted 12 droplets at the location given by eqn.~\ref{eq:am_vid}. Each droplet was placed with a random initial location, i.e. at pixel location given as $(y^i(t=0)$,$x^i(t=0))$. These images were then sequentially stacked to prepare a video of continuously moving droplets on predefined paths. In Fig.~\ref{fig:sythetic_track} we show one instance of the droplet identification and tracking process. See video6.avi for the output of the developed application.

\bibliography{Ref}

\end{document}